\documentclass[twocolumn,letter]{jpsj2} 
\title{Multigap Superconductivity in Y$_2$C$_3$: A $^{13}$C-NMR Study}
\author{A. \textsc{Harada}$^{1}$\thanks{E-mail address: aharada@nmr.mp.es.osaka-u.ac.jp}, S. \textsc{Akutagawa}$^{2}$, Y. \textsc{Miyamichi}$^{1}$, H. \textsc{Mukuda}$^{1}$, Y. \textsc{Kitaoka}$^{1}$ and J. \textsc{Akimitsu}$^{2}$}

\inst{$^{1}$Department of Materials Engineering Science, Graduate School of Engineering Science, Osaka University, Toyonaka, Osaka 560-8531, Japan \\
$^{2}$Department of Physics and Mathematics, Aoyama-Gakuin University, Sagamihara, Kanagawa 229-8558, Japan\\ }

\recdate{\today}

\abst{We report on the superconducting (SC) properties of Y$_2$C$_3$ with a relatively high transition temperature $T_{\rm c}=15.7$\,K investigated by $^{13}$C nuclear-magnetic-resonance (NMR) measurements under a magnetic field. The  $^{13}$C Knight shift has revealed a significant decrease below $T_{\rm c}$, suggesting a spin-singlet superconductivity.   
From an analysis of the temperature dependence of the nuclear spin-lattice relaxation rate $1/T_1$ in the SC state, Y$_2$C$_3$ is demonstrated to be a multigap superconductor that exhibits a large gap $2\Delta/k_{\rm B}T_{\rm c}=5$ at the main band  and a small gap $2\Delta/k_{\rm B}T_{\rm c}=2$ at other bands. These results have revealed that Y$_2$C$_3$ is a unique multigap s-wave superconductor similar to MgB$_2$. }

\kword{lanthanoid carbide, metal, superconductivity, multigap, NMR, Y$_2$C$_3$}

\newpage
\begin{document}
\maketitle

The discovery of superconductivity in MgB$_2$, which exhibits a high superconducting (SC) transition temperature $T_{\rm c}\sim 40$\,K, has attracted much interest.\cite{Nagamatsu} Motivated by this discovery, intensive effort is devoted to the search for a new high-$T_{\rm c}$ material in a similar system that contains light elements B and C. 
Meanwhile, Amano $et~al$.~reported that Y$_2$C$_3$ prepared under high pressure ($\sim 5$\,GPa) is a superconductor with a relatively high $T_{\rm c}\sim 18$\,K,\cite{Amano} although the superconductivity in this compound was already reported to emerge at $T_{\rm c}\sim 6-11$\,K.\cite{Krupka}  As for SC characteristics, the specific heat measurements on the newly synthesized high-purity samples of Y$_2$C$_3$ have revealed that a gap size of the respective samples with $T_{\rm c}=11.6$, 13.6, and 15.2\,K increases as $2\Delta/k_{\rm B}T_{\rm c}$=3.6, 4.1, and 4.4.\cite{Akutagawa} 
This result raises a question why $T_{\rm c}$ and $2\Delta/k_{\rm B}T_{\rm c}$ vary significantly depending on sintering conditions.  Further systematic experiments are required to gain deep insight into the SC characteristics of this compound.  

\begin{figure}[h]
\centering
\includegraphics[width=7.0cm]{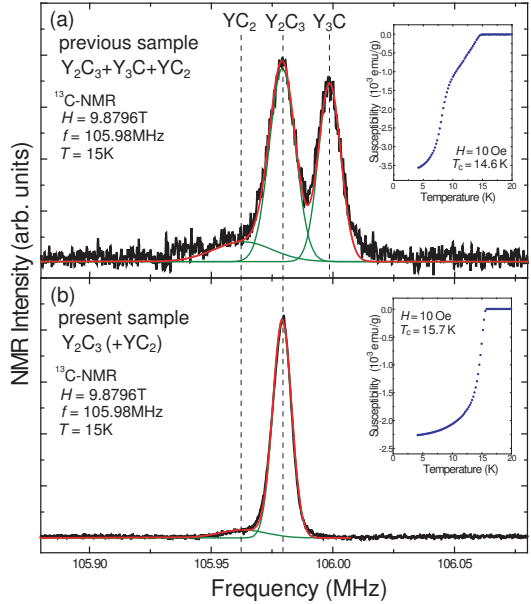}
\caption[]{(Color online) $^{13}$C-NMR spectra of Y$_2$C$_3$  for (a) the previous sample reported in ref.~2 and  (b) the present sample in the normal state at $T=15$\,K and $H=9.85$\,T. Note that the spectra for the previous sample are composed of NMR signals arising from Y$_3$C, YC$_2$, and Y$_2$C$_3$, demonstrating the contamination of extrinsic multiphases, whereas the spectra for the present sample nearly consist of a single peak from Y$_2$C$_3$ with a small contamination of YC$_2$. The insets of both figures show the $T$ dependence of SC diamagnetic susceptibility down to 4.2\,K.}
\label{spectra}
\end{figure}

Y$_2$C$_3$ crystallizes in the cubic Pu$_2$C$_3$-type structure (space group $I\bar{4}3d$) without an inversion center, consisting of the dimers of carbon atoms. The SC properties of the sample previously reported by Amano $et~al$. did not exhibit a single SC transition as seen in the inset of Fig.~\ref{spectra}(a), pointing to a contamination of extrinsic multiple phases.\cite{Amano}  Recently, Akutagawa $et~al$.~have succeeded in preparing a single phase of Y$_2$C$_3$, which enables us to extract intrinsic electronic and SC properties in Y$_2$C$_3$. 
From the specific-heat measurements of this sample, the Sommerfeld coefficient $\gamma \sim 6.3$\,mJ/mol$\cdot$K$^2$ and Debye temperature $\theta_{\rm D}\sim 530$\,K were estimated for the sample with $T_{\rm c}=15.2$\,K.\cite{Akutagawa} This result suggests that its high Debye temperature makes $T_{\rm c}$ relatively high despite its small Sommerfeld coefficient.  As in MgB$_2$, the light-element constituent like boron and carbon plays a vital role in enhancing $T_{\rm c}$ in general. In the SC state, the temperature ($T$) dependence of the specific heat exhibits an exponential decrease with $2\Delta$/$k_{\rm B}T_{\rm c}=4.4$ upon cooling well below $T_{\rm c}$, suggesting a strong-coupling isotropic superconductivity. From an other context, it is noteworthy that a novel SC nature for CePt$_3$Si and Li$_2$Pt$_3$B without inversion symmetry is a recent interesting topic because the admixture of spin-singlet and spin-triplet SC state is shown to emerge due to the spin-orbit coupling.\cite{Yogi,Yuan} Likewise, determining the 
order-parameter symmetry and a detailed gap structure is an underlying issue in the newly synthesized high-quality Y$_2$C$_3$ without inversion symmetry.    

In this letter, we report on the SC order-parameter symmetry and gap structure of Y$_2$C$_3$ with a relatively high $T_{\rm c}=15.7$\,K ($H=0$) via $^{13}$C nuclear-magnetic-resonance (NMR) measurements under a magnetic field.
Y$_2$C$_3$ was synthesized by arc melting and high pressure.\cite{Akutagawa} The sample was confirmed to nearly consist of a single phase by X-ray diffraction analyses, with the formation of a primitive Pu$_2$C$_3$-type structure. The polycrystalline sample for $^{13}$C-NMR measurement was slightly enriched with $^{13}$C ($^{12}$C~:~$^{13}$C~=~9~:~1) in order to improve the NMR signal-to-noise ratio. $T_{\rm c}=15.7$ and 12.2\,K were determined  by ac-susceptibility measurements at $H=0$ and $9.85$\,T, respectively. The NMR experiment was performed by the conventional spin-echo method at $H=9.85$\,T in the $T$ range of $1.8-70$\,K. 

Figures~\ref{spectra}(a) and \ref{spectra}(b) show the $^{13}$C-NMR spectra of Y$_2$C$_3$ for the previous sample reported in ref.~2 and  the present sample, respectively, in the normal state at $T=15$\,K and $H=9.85$\,T. Note that the spectra for the previous sample are composed of $^{13}$C-NMR signals arising from Y$_3$C, YC$_2$, and Y$_2$C$_3$, demonstrating the contamination of extrinsic multiphases, whereas the spectra for the present sample consist of a nearly single peak from Y$_2$C$_3$ with a small contamination of YC$_2$. 
The NMR intensity for each phase coincides with the x-ray intensity as expected. The full width at half maximum (FWHM) in the $^{13}$C-NMR spectrum of Y$_2$C$_3$ is as small as 8\,kHz, ensuring good sample quality.
Actually, a single SC transition at $T_{\rm c}=15.7$\,K was corroborated by the susceptibility measurement at $H=10$\,Oe as seen in the inset of Fig.~\ref{spectra}(b). 

\begin{figure}[h]
\centering
\includegraphics[width=7.0cm]{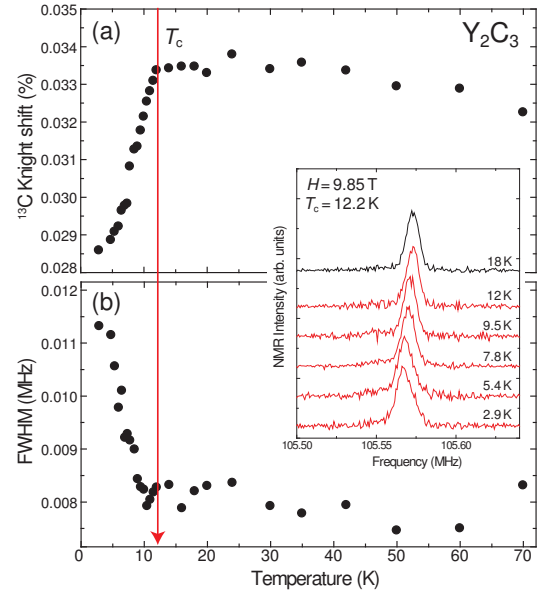}
\caption[]{(Color online) $T$ dependences of (a) $^{13}$C Knight shift and (b) full width at half maximum (FWHM) of $^{13}$C-NMR spectrum in Y$_2$C$_3$ at $H=9.85$\,T. $T_{\rm c}=12.2$\,K is determined by the ac-susceptibility measurement at $H=9.85$\,T as shown by an arrow pointing downward. The inset shows the NMR spectra at $T=2.9$, 5.4, 7.8, 9.5, 12, and 18\,K, respectively.}
\label{shift}
\end{figure}

Figure~\ref{shift}(a) shows the $T$ dependence of $^{13}$C Knight shift (KS) in Y$_2$C$_3$, which is determined relative to the resonance frequency of tetramethylsilane (TMS) as a reference substance ($K{\rm [TMS]} \sim 0\,{\rm ppm}$). 
A clear decrease in KS and an increase in FWHM below $T_{\rm c}$ are indicated in Figs.~\ref{shift}(a) and \ref{shift}(b), which are derived from the $T$ dependence of NMR spectra as shown in the inset of Fig.~\ref{shift}.   
%
In intermediate fields ($H_{\rm c1}\ll H \ll H_{\rm c2}$) where the vortices form a dense lattice, we estimate coherence length and the distance between the vortices to be $d\sim 160$\,\AA, and $\xi\sim 34$\,\AA, respectively.\cite{Keisan}
As a result, a diamagnetic field is led to be $H_{\rm dia}\sim -0.3$\,Oe using the relation $H_{\rm dia}=-H_{\rm c1}{\rm ln}(\beta e^{-1/2} d/\xi)/{\rm ln}\kappa$,\cite{Gennes,Keisan} ($\kappa=\lambda/\xi$) and hence a diamagnetic shift is obtained as $K_{\rm dia}\sim 3 \times 10^{-4}$\,\% at $H$=9.85\,T.  
Here, we used $H_{\rm c1}=3.3$\,mT,\cite{Akutagawa} the London penetration depth $\lambda=4470$\,\AA,\cite{Akutagawa} $\beta=0.381$ for the case of triangular lattice,\cite{Gennes} $d\sim 160$\,\AA, and $\xi\sim 34$\,\AA.
Thus, the estimated value of $K_{\rm dia}$ is one order of magnitude smaller than the decrease in KS observed below $T_{\rm c}$, demonstrating that the decrease in KS is due to the reduction of spin susceptibility associated with the onset of the spin-singlet SC state. If the spin susceptibility was assumed to vanish at low $T$ due to the formation of spin-singlet Cooper pairing, the orbital and spin part of KS are tentatively estimated to be $K_{\rm orb}\sim 0.028$\,\% and $K_{\rm s}\sim 0.005$\,\%, respectively. 
A possible cause for the increase in FWHM may be due to an inhomogeneous distribution of vortex lattices, which eventually makes either $d$ or $\lambda$ distribute. Further systematic NMR measurements at low $H$ are required for inspecting a structure of vortex lattices. 
\begin{figure}[h]
\centering
\includegraphics[width=7.0cm]{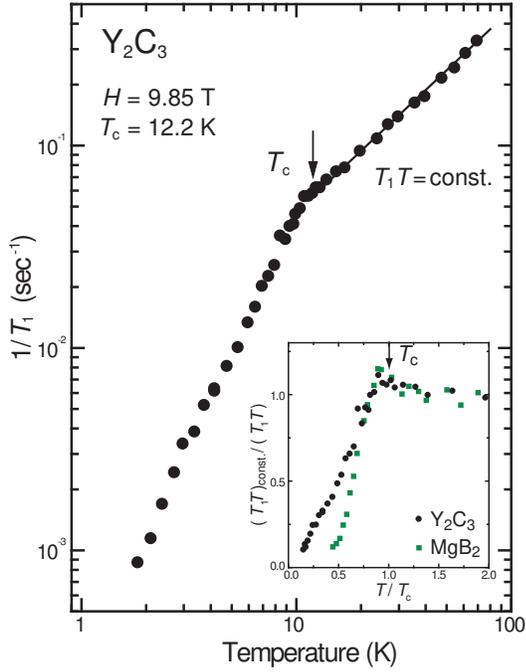}
\caption[]{(Color online) $T$ dependence of 1/$T_1$ for Y$_2$C$_3$ with $T_{\rm c}= 12.2$\,K at $H\sim 9.85$\,T. A tiny coherence peak just below $T_{\rm c}$ is observed for Y$_2$C$_3$ as in MgB$_2$ as shown in the inset. The inset shows the $(T_1T)_{\rm const.}$/$(T_1T)$ vs $T/T_{\rm c}$ curve for Y$_2$C$_3$ (solid circles) at $H=9.85$\,T and for MgB$_2$ (solid squares) with $T_{\rm c}= 29$\,K at $H=4.4$\,T.\cite{Kotegawa} Here, $(T_1T)_{\rm const.}$ denotes constant values in normal state.}
\label{1/T1vsT}
\end{figure}

\begin{figure}[hb]
\centering
\includegraphics[width=7.5cm]{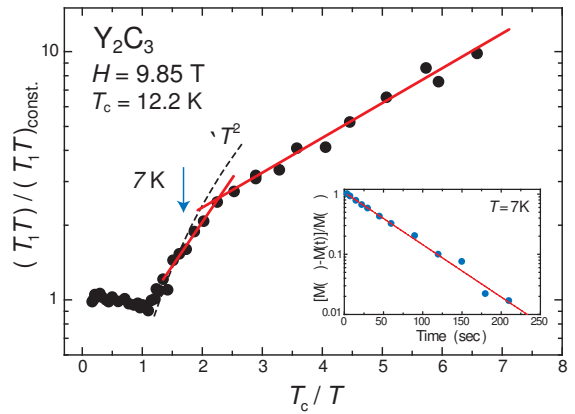}
\caption[]{(Color online) Arrhenius plot of $(T_1T)/(T_1T)_{\rm const.}$ vs $T_{\rm c}/T$ with $T_{\rm c}=12.2$\,K at $H=9.85$\,T. From the respective slopes in the $T$ range 5\,K $-T_{\rm c}=12.2$ K and at temperatures lower than 5\,K, gap sizes are estimated as $2\Delta/k_{\rm B}T_{\rm c}\sim 5$ and 2. The inset shows a simple exponential recovery curve of nuclear magnetization given by the relation $[M(\infty)-M(t)]/M(\infty)=\exp(-t/T_1)$. Here, $M(t)$ and $M(\infty)$ are the nuclear magnetizations at a time $t$ after the saturation pulse and at the thermal equilibrium condition, respectively. Even in the SC state at $T=7$\,K, note that $1/T_1$ for $^{13}$C with nuclear spin $I=1/2$ is uniquely determined (see the text).}
\label{T1TvsT}
\end{figure}

Next, we deal with the $T$ dependence of nuclear spin-lattice relaxation rate $(1/T_1)$ in order to clarify the gap structure. The $1/T_1$ for $^{13}$C with nuclear spin $I=1/2$ is uniquely determined from a simple exponential recovery curve of nuclear magnetization given by the relation $[M(\infty)-M(t)]/M(\infty)=\exp(-t/T_1)$. Here, $M(t)$ and $M(\infty)$ are the nuclear magnetizations at a time $t$ after the saturation pulse and at the thermal equilibrium condition, respectively. Figure~\ref{1/T1vsT} presents the $T$ dependence of $1/T_1$ at $H=9.85$\,T. In the normal state, the law $T_1T={\rm const.}$ is valid down to $T_{\rm c}$. In the SC state, $1/T_1$ is also precisely measured from a simple exponential recovery curve such as $[M(\infty)-M(t)]/M(\infty)=\exp(-t/T_1)$, as seen in the inset of Fig.~\ref{T1TvsT}(a). This is because the possible contribution to $1/T_1$ arising from normal vortex cores is very small, if any, when $\xi\sim 34$\,\AA $<<$ $d\sim 160$\,\AA.  

The inset in Fig.~\ref{1/T1vsT} shows $(T_1T)_{\rm const.}$/$(T_1T)$ vs $T/T_{\rm c}$ for Y$_2$C$_3$ at $H=9.85$\,T and for MgB$_2$ with $T_{\rm c}= 29$\,K at $H=4.4$\,T.\cite{Kotegawa} Here, $(T_1T)_{\rm const.}$ denotes constant values in normal state. In the SC state, we note that a tiny coherence peak is observed in $1/T_1$ just below $T_{\rm c}$ for Y$_2$C$_3$ as in MgB$_2$. This is indicative of a full gap opening in Y$_2$C$_3$ as in MgB$_2$. A reason why the coherence peak is depressed in these compounds may be due to a strong electron-phonon coupling that causes the large life time broadening of quasiparticles induced by thermally excited phonons as reported in ref.~10. In fact, the strong-coupling BCS superconductor such as TlMo$_6$Se$_{7.5}$ does not show a clear coherence peak.\cite{Ohsugi}  Note that the $T$ dependence of $1/T_1$ well below $T_{\rm c}$  does not exhibit a simple exponential decrease, but seems to have a kink at around $T=5$\,K. In order to gain further insight into this unique and relevant relaxation behavior with a possible gap structure in the SC state for Y$_2$C$_3$, we present in Fig.~\ref{T1TvsT} the Arrhenius plot of $(T_1T)/(T_1T)_{\rm const.}$ vs $T_{\rm c}/T$ with $T_{\rm c}=12.2$\,K at $H=9.85$\,T.
It is evident that a power-law behavior in $1/T_1$ such as 1/$T_1T\propto T^2$ (see the dashed line in the figure) is not valid at all. Instead, when noting that a line in this plot corresponds to an exponential $T$ dependence in $1/T_1T$, it is  supposed that a large full gap seemingly opens in a high-temperature regime in the SC state, but low-lying quasiparticle excitations in a low-temperature regime are dominated by the presence of a small full gap. In fact, from respective slopes in this plot in the $T$ range of 5\,K-$T_{\rm c}=12.2$\,K and at temperatures lower than 5\,K, gap sizes are estimated to be $2\Delta/k_{\rm B}T_{\rm c}\sim$ 5 and 2, which suggests that two kinds of SC energy gaps exist, namely, multigap superconductivity takes place in Y$_2$C$_3$. We stress that this novel relaxation behavior in the SC state for Y$_2$C$_3$ is not due to some inhomogeneous effect originating from the presence of vortex cores and/or a distribution of $T_{\rm c}$ because $1/T_1$ is uniquely determined from the simple exponential recovery curve of nuclear magnetization as shown in the inset of Fig.~\ref{T1TvsT}.  
\begin{figure}[h]
\centering
\includegraphics[width=7.5cm]{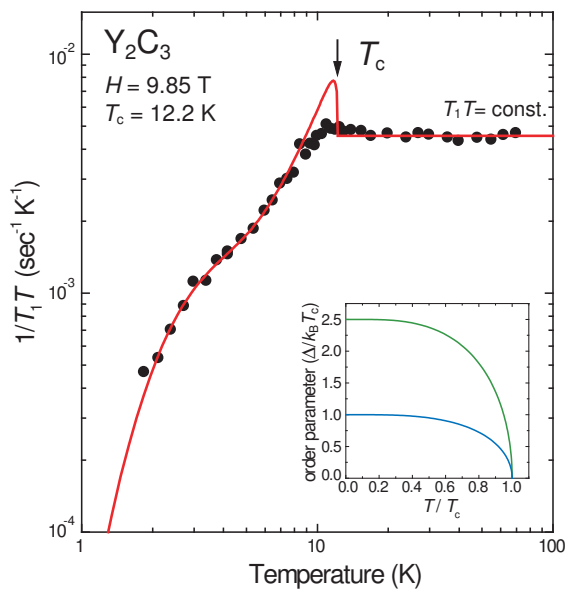}
\caption[]{(Color online) $T$ dependence of $1/T_1T$ for Y$_2$C$_3$ with $T_{\rm c}= 12.2$\,K at $H=9.85$\,T. 
A phenomenological multigap model for nodeless superconductivity is applied to understand the novel relaxation behavior in the SC state. The solid curve is a theoretical curve based on the multigap model with $\beta=0.75$ and $2\Delta_{\beta}/k_{\rm B}T_{\rm c}=5$ for the main band, and $\alpha=0.25$ and $2\Delta_{\alpha}/k_{\rm B}T_{\rm c}=2$ for other band (see the text). The inset shows the $T$ dependences of the order parameters $\Delta_{\alpha}$ and $\Delta_{\beta}$. }
\label{1/T11vst}
\end{figure}

Here, we apply a phenomenological multigap model for nodeless superconductivity to understand the novel relaxation behavior in the SC state. Figure~\ref{1/T11vst} shows the $T$ dependence of $1/T_1T$. 
In such a model, $T_1(T_{\rm c})/T_1(T)$ is expressed as
\[\displaystyle \frac{T_1(T_{\rm c})}{T_1(T)}=\frac{\alpha^2}{\alpha^2+\beta^2}\,\frac{1}{T_1}(\Delta_{\alpha})+\frac{\beta^2}{\alpha^2+\beta^2}\,\frac{1}{T_1}(\Delta_{\beta}),\]
where $\alpha$ and $\beta$  are defined as the respective fractions of $N({\rm E_{\rm F}})\times A_{\rm hf}$ with SC gap $\Delta_{\alpha}$ and $\Delta_{\beta}$ and $\alpha+\beta=1$. $N({\rm E_{\rm F}})$ and $A_{\rm hf}$ are the density of states (DOS) at the Fermi level and the hyperfine coupling constant, respectively. Here,
\[\displaystyle \frac{1}{T_1}(\Delta)=\frac{2}{k_{\rm B}T_{\rm c}}\int^{\infty}_0 dE \, [N_{\rm s}^2(E)+ M_{\rm s}^2(E)] f(E)[1-f(E)],\]
where $N_{\rm s}(E)$ is the DOS, $M_{\rm s}(E)$ is the anomalous DOS originating from the coherence effect inherent to a spin-singlet SC state and $f(E)$ is the Fermi distribution function.\cite{Hebel1}
Note that $N_{\rm s}(E)$ and $M_{\rm s}(E)$ are averaged over an energy broadening function assuming a rectangle shape with a width 2$\delta$ and a height 1/2$\delta$.\cite{Hebel2} We use $\delta/\Delta(0)=0.3$ in the calculation. 
A theoretical curve based on the multigap model is actually in good agreement with the experiment using $\beta=0.75$ and $2\Delta_{\beta}/k_{\rm B}T_{\rm c}=5$ for the main band, and $\alpha=0.25$ and $2\Delta_{\alpha}/k_{\rm B}T_{\rm c}=2$ for other bands as shown by the solid curve in Fig.~\ref{1/T11vst}. It is notable that the large gap at the dominant Fermi surface is larger than the weak-coupling BCS value of $2\Delta/k_{\rm B}T_{\rm c}=3.5$, indicating a strong electron-phonon coupling and being consistent with the specific-heat result.\cite{Akutagawa} 
The present $^{13}$C-NMR has revealed that the superconductivity in Y$_2$C$_3$ is characterized by a large gap at the main Fermi surface and a small gap at others. This may be consistent with the band calculation which shows the presence of Fermi surfaces consisting of three dimensional multisheets due to the hybridization between Y-$d$ derived states and antibonding C-dimers derived $p$-states.\cite{Singh} We should pay attention to the relationship between the multigap and $T_{\rm c}$. Although Y$_2$C$_3$ is a superconductor  with no inversion symmetry, the present experiments have revealed that this noncentrosymmetric compound is a 
spin-singlet superconductor with full gaps at all the Fermi surfaces, and hence rules out the possibility of the admixture of spin-triplet order parameter which is the recent underlying topic for the superconductors with no inversion symmetry.    

In conclusion, the superconducting properties of Y$_2$C$_3$ with a relatively high transition temperature $T_{\rm c}=15.7$\,K have been investigated using the $^{13}$C nuclear-magnetic-resonance (NMR) method under a magnetic field. The Knight shift has revealed a significant decrease below $T_{\rm c}$, suggesting the spin-singlet superconductivity.  
The nuclear spin-lattice relaxation study in the SC state has revealed that Y$_2$C$_3$ is a multigap superconductor that exhibits a large gap $2\Delta/k_{\rm B}T_{\rm c}=5$ at the main band and a small gap $2\Delta/k_{\rm B}T_{\rm c}=2$ at others.
These results have revealed that Y$_2$C$_3$ is a unique multigap s-wave superconductor similar to MgB$_2$.

The authors would like to thank H.~Kotegawa, M.~Yogi, and H.~Tou for fruitful discussions and comments, and N.~Terasaki for experimental support.
This work was supported by a Grant-in-Aid for Creative Scientific Researchi15GS0213), MEXT and the 21st Century COE Program supported by the Japan Society for the Promotion of Science.



\begin{thebibliography}{99}
\bibitem{Keisan}
We use $H_{\rm c2}=-0.691(dH_{c2}/dT)T_{\rm c}$ and $\xi^2=\phi_0/2\pi H_{\rm c2}$ to get the value of $\xi$.
$d$ is estimated using $d^2=2\phi_0/\sqrt{3} B$ (triangular lattice). $\phi_0$ is the flux quantum.
\end{thebibliography}
\end{document}